\newif{\ifjournal}
  \journalfalse

\ifjournal
  \documentclass[]{aa}
  \usepackage{graphicx,times,amssymb,natbib}
  \renewcommand{\d}{\mathrm{d}}
  \authorrunning{K. Dolag et al.}
  \titlerunning{Numerical study of halo concentrations in dark-energy
    cosmologies}
\else
  \documentclass{paper}
\fi

\begin{document}

\title{Numerical study of halo concentrations in dark-energy
  cosmologies}
\ifjournal
  \author{Klaus Dolag\inst{1}, Matthias
    Bartelmann\inst{2}, Francesca Perrotta\inst{3,4}, Carlo
    Baccigalupi\inst{3,4},\\ Lauro Moscardini\inst{5}, Massimo
    Meneghetti\inst{1} and Giuseppe Tormen\inst{1}
  \institute
   {$^1$ Dipartimento di Astronomia, Universit\`a di Padova, Italy\\
    $^2$ Max-Planck-Institut f\"ur Astrophysik, P.O.~Box 1317,
    D--85741 Garching, Germany\\
    $^3$ SISSA, Trieste, Italy\\
    $^4$ Berkeley, USA\\
    $^5$ Dipartimento di Astronomia, Universit\`a di Bologna, Italy}}
\else
  \author{Klaus Dolag$^1$, Matthias
    Bartelmann$^2$\footnote{\emph{present address}: ITA, Universit\"at
    Heidelberg, Tiergartenstr.~15, D--69121 Heidelberg, Germany},
    Francesca Perrotta$^{3,4}$, Carlo Baccigalupi$^{3,4}$,\\ Lauro
    Moscardini$^5$, Massimo Meneghetti$^1$ and Giuseppe Tormen$^1$\\
    $^1$ Dipartimento di Astronomia, Universit\`a di Padova, Italy\\
    $^2$ Max-Planck-Institut f\"ur Astrophysik, P.O.~Box 1317,
    D--85741 Garching, Germany\\
    $^3$ SISSA, Trieste, Italy\\
    $^4$ Berkeley, USA\\
    $^5$ Dipartimento di Astronomia, Universit\`a di Bologna, Italy}
\fi
\date{\emph{Astronomy \& Astrophysics, submitted}}

\newcommand{\abstext}
 {We study the concentration parameters, their mass dependence and
  redshift evolution, of dark-matter halos in different dark-energy
  cosmologies with constant and time-variable equation of state, and
  compare them with ``standard'' $\Lambda$CDM and OCDM models. We find
  that previously proposed algorithms for predicting halo
  concentrations can be well adapted to dark-energy models. When
  centred on the analytically expected values, halo concentrations
  show a log-normal distribution with a uniform standard deviation of
  $\sim0.2$. The dependence of averaged halo concentrations on mass
  and redshift permits a simple fit of the form
  $(1+z)\,c=c_0\,(M/M_0)^\alpha$, with $\alpha\approx-0.1$
  throughout. We find that the cluster concentration depends on the
  dark energy equation of state at the cluster formation redshift
  $z_\mathrm{coll}$ through the linear growth factor
  $D_+(z_\mathrm{coll})$. As a simple correction accounting for
  dark-energy cosmologies, we propose scaling $c_0$ from $\Lambda$CDM
  with the ratio of linear growth factors, $c_0\rightarrow
  c_0\,D_+(z_\mathrm{coll})/D_{+,\Lambda\mathrm{CDM}}(z_\mathrm{coll})$.}

\ifjournal
  \abstract{\abstext}
\else
  \begin{abstract}\abstext\end{abstract}
\fi

\maketitle

\section{Introduction}

The properties of dark-matter halos have been shown in many earlier
studies to depend in characteristic ways on their formation
history. The dark-matter density profile consistently found in
numerical simulations by \citep{NA96.1,NA97.1} has a radial scale,
$r_\mathrm{s}$, which is smaller than the virial radius
$r_\mathrm{vir}$ by the concentration parameter
$c=r_\mathrm{vir}/r_\mathrm{s}$. Although there is an ongoing
discussion as to the exact density-profile slope in halo cores, there
is agreement that the density profile steepens substantially going
outward across the scale radius.

The central density in halo cores was found to reflect the mean
density of the Universe at the time of halo formation. Since halos of
increasing mass form at increasingly late cosmic epochs in hierarchical
models of structure formation, the concentration parameter of halos
at fixed redshift decreases with mass. For halos of fixed mass, the
concentration parameter increases with decreasing redshift because the
background density drops.

Structure growth in the Universe is thus reflected by the core
densities, or concentrations, of dark-matter halos. Haloes thus
establish a connection between principally measurable quantities, like
their central densities, and the cosmological framework model because
the latter dictates how structures form.

We have shown in an earlier paper \citep{BA02.1} that halos of given
mass and redshift are predicted to have higher concentrations in
dark-energy cosmologies than in models with cosmological constant
because they tend to form earlier and thus acquire a higher core
density. For that study, we used analytic algorithms for predicting
halo properties which were assumed, but never shown, to hold true in
dark-energy cosmologies. In this paper, we describe numerical
simulations undertaken with the specific aim of studying how halo
properties change in dark-energy models, and whether the analytic
algorithms proposed for other cosmologies can be adopted or adapted to
dark-energy models.

Klypin et al. \citep{KL03.1} also studied the properties of clusters
simulated in the framework of cosmological models with dynamical dark
energy. However, their cosmological models differ from ours for the
choice of many parameters. Moreover, we will analyze here a larger
sample of individually simulated clusters, covering a completely
different range of halo masses.

We describe the dark-energy models used in Sect.~2 and our numerical
simulations in Sect.~3. Sect.~4 details the determination of halo
concentrations, their comparison with analytic expectations, and their
statistical properties. We identify an interesting dependence between
average halo concentrations and the density of dark energy at halo
formation in Sect.~5, and conclude with a summary in Sect.~6.

\section{Dark-energy models\label{dem}}

In this paper, we study and compare results obtained for the following
cosmological models: an open Cold Dark Matter (OCDM) and four flat
dark-energy cosmogonies. The latter are a cosmological constant
($\Lambda$CDM) model, a dark-energy model with constant equation of
state (DECDM), and two quintessence models, one with inverse power-law
Ratra-Peebles potential \citep[RP, see][ and references
therein]{PE03.1} and one with SUGRA potential \citep[SUGRA, see][ and
references therein]{BR00.2}.

In all cases, the matter density parameter today is $\Omega_0=0.3$. In
the flat cosmologies, the remaining $70\%$ of the critical density is
assigned to the dark energy at present. The remaining cosmological
parameters are $h=0.7$, $\Omega_\mathrm{b}\,h^{2}=0.022$, a Gaussian
density fluctuations with scale-invariant power spectrum, and no
gravitational waves.

Particularly important for us is the normalisation of the perturbation
power spectrum, which we set by defining the \emph{rms} density
fluctuation level within spheres of $8\,h^{-1}\,\mathrm{Mpc}$ radius,
$\sigma_8$. In this respect, we follow two approaches, which are
normalising the perturbation amplitude either on large scales with the
observed Cosmic Microwave Background (CMB) anisotropies
\citep[e.g.][]{BE03.3}, or on small scales using the observed cluster
abundance. In the second case, we choose $\sigma_8=0.9$ in all the
models, while the $\sigma_8$ derived from the CMB is generally
slightly smaller because of the Integrated Sachs Wolfe (ISW) effect
affecting the large scale CMB anisotropies in the cosmologies we
consider \citep[see][and references therein]{BA02.1}: we take $0.86$,
$0.82$ and $0.76$ for DECDM, RP and SUGRA, respectively. These numbers
as well as all the inputs from the linear evolution of cosmological
perturbations to the $N$-body procedure described later are computed
using our dark energy oriented cosmological code \citep{PE99.4}, based
on CMBfast \citep{SE96.1}.

We briefly describe now the dark energy cosmologies we will adopt. The
equation of state $w$ is a key parameter, describing the ratio between
the dark energy pressure and energy density, and it must be negative
in order to effectuate cosmic acceleration today. In the limit of $w$
being constant, the dark energy density evolves in redshift
proportionally to $(1+z)^{3(1+w)}$. That is the case for both the
$\Lambda$CDM model, whose $w=-1$ yields a constant dark energy
density, and the DECDM model, which we choose to have $w=-0.6$. Note
that the latter case is an effective scenario in the sense that it is
obtained simply by replacing the cosmological-constant equation of
state, assuming it to be constant and neglecting any other dark energy
features such as its cosmological perturbations. Also, it is useful to
note that from the point of view of the Friedmann equation, the
curvature term in the OCDM model behaves as a dark energy component
with constant $w=-1/3$ \citep[see][ and references therein]{BA02.1}.

The dark energy is consistently described by means of the quintessence
scalar field $\phi$. The dynamics of its unperturbed value and of the
linear fluctuation $\delta\phi$ obey the Klein-Gordon equation
\begin{equation}
  \Box{\phi}+V'=0\quad,\quad\delta(\Box{\phi})+V''\delta\phi=0\;,
\label{KG}
\end{equation}
where $\Box$ is the d'Alambert operator in a
Friedmann-Robertson-Walker (FRW) cosmology, and $V(\phi)$ is the
quintessence potential.

The RP and SUGRA potentials are given by
\begin{equation}
  V_\mathrm{RP}=\frac{M^{4+\alpha_Q}}{\phi^{\alpha_Q}}\quad,\quad
  V_\mathrm{SUGRA}=\frac{M^{4+\alpha_Q}}{\phi^{\alpha_Q}}\cdot
  \exp{(4\pi G\phi^2)}\;,
\label{RPSUGRA}
\end{equation}
respectively. The exponential in $V_\mathrm{SUGRA}$ comes from
super-gravity corrections \citep{BR00.2} and induces large changes in
the trajectories with respect to the RP case, as we explain now.

The attractor properties of the tracking trajectories in the
quintessence scenarios allow to reach the present field value
$\phi_0$, of the order of the Planck mass $M_\mathrm{Planck}$,
starting from a wide set of initial conditions for $\phi$ and
$\dot{\phi}$, with the only relevant condition that $\phi_i\ll
M_\mathrm{Planck}$. For both RP and SUGRA, the tracking trajectories
are well defined until the quintessence is subdominant compared to the
matter density, yielding a constant equation of state obeying the
simple relation
\begin{equation}
  w=-\frac{2}{2+\alpha_Q}\;.
\label{track}
\end{equation}
In the RP scenario, if we require the \emph{present} equation of state
$w_0$ to respect the current constraints \citep[see][ and references
therein]{BE03.3}, say $w_0\le-0.8$, the exponent must be in the range
$0\le\alpha_{Q}\le0.5$, yielding a shallow potential shape. At
present, the tracking regime is abandoned, but the shallow potential
does not permit $w_0$ to be far from the tracking solution of
Eq.~(\ref{track}), the deviation being at the $10\%$ level.

The SUGRA exponential correction flattens the potential shape
noticeably at $\phi\simeq M_\mathrm{Planck}$, i.e.~at the \emph{end}
of the tracking trajectory. This means that a given equation of state
at present is obtained for noticeably higher values of $\alpha_Q$ than
for RP, i.e.~the dark energy dynamics and thus the cosmological
expansion rate as a function of redshift are generally much different
in the two cases. 

\begin{figure}
  \includegraphics[width=\hsize]{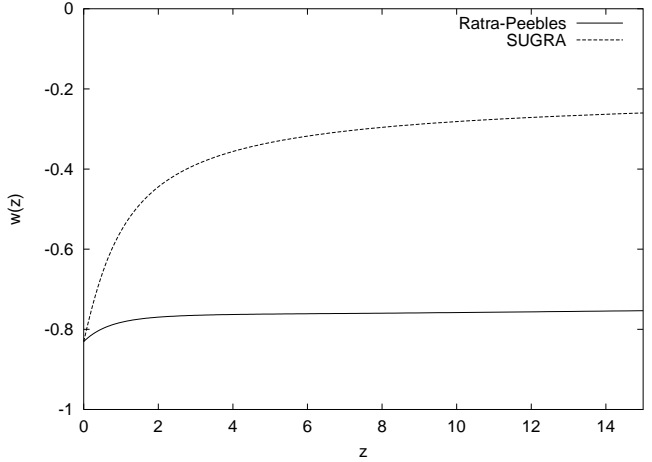}
\caption{Equation of state for the RP and SUGRA quintessence models
  studied in this work.}
\label{w_z_RP_SUGRA}
\end{figure}

Thus, we consider two quintessence models here having the same
equation of state at the present epoch, but markedly different ones in
the past. Specifically, they are tracking quintessence scenarios
obtained out of the RP and SUGRA potentials, both having $w_0=-0.83$
today, with $\alpha_{Q}=-0.6$ and $\alpha_{Q}=-6.7$,
respectively. According to Eq.~(\ref{track}), this yields $w_{\rm
SUGRA}\simeq -0.23$ and $w_{\rm RP}\simeq -0.77$ in the tracking
regime, well representing the diversity that dark energy cosmologies
can have in the past, even if they reproduce the same amount of cosmic
acceleration today. The redshift behaviour of $w$ in the two cases is
illustrated in Fig.~\ref{w_z_RP_SUGRA}.

Note that if $w$ depends on $z$, the dark energy density evolves as
\begin{equation}
  \rho_\mathrm{DE}(z)=\rho_\mathrm{DE,0}\,
  \exp\left[3\int_0^z\d z'\frac{1+w(z')}{1+z'}\right]\;,
\label{e:omegade}
\end{equation}
which is easily verified solving the energy conservation equation
$\dot{\rho}+3\,H\,(\rho+p)=0$.

The cosmological scenarios we study here are mutually very
different. The differences are relevant even between the flat dark
energy models: for DECDM, the equation of state is far from $-1$,
while RP and SUGRA have the same equation of state today, but differ
strongly in the past, as we emphasised above. These pronounced
differences allow us to recognise their imprint in the properties of
the dark non-linear halos which we study next. In this respect, we
obtain the most interesting results when the cosmologies above are
compared by keeping everything else fixed, including $\sigma_8$.

\section{Numerical simulations}

\subsection{The simulation code}

We adapted the cosmological code GADGET (\citealt{SP01.1}) for
simulating a set of 17 galaxy clusters in various dark-energy
cosmologies with constant and time-variable equations of state. This
section describes the code used, the modifications required for
dark-energy models, and the initial conditions we set up.

In total we performed 136 cluster simulations. The simulations were 
performed in parallel on 8 CPUs
on an IBM-SP4 located at the CINECA Super Computing Centre in Bologna,
Italy.

\subsubsection{GADGET}

The cosmological code GADGET (Springel et al. 2001b) is well tested
and used for simulating a wide range of cosmological scenarios from
the interaction of galaxies \citep{SP00.1} to large-scale structure
formation \citep{SP01.3}. GADGET is based on a tree-SPH code using
co-moving coordinates. The new version (P-GADGET-2, kindly provided by
Volker Springel) allows the computation of long-range forces with a
particle-mesh (PM) algorithm, with the tree algorithm supplying
short-range gravitational interactions only. This ``TreePM'' method
can substantially speed up the computation while maintaining the large
dynamic range and flexibility of the tree algorithm. It also differs
in the choice of internal variables for time and velocity, in its
time-stepping algorithm and in its parallelisation strategy (see
Springel 2003, in preparation). The modifications necessary for
performing simulations within cosmologies with dark-energy within
P-GADGET-2 do not differ in principle from the modifications necessary
for the earlier version of GADGET, but appear in slightly different
places due to the changes in design of the code.

\subsubsection{Including dark-energy in GADGET}

Within the co-moving coordinate scheme of GADGET, the only place where
the dark energy has to be taken care of is the calculation of the
Hubble function, or some combination of the Hubble function with the
scale factor $a$. This is used when ever a conversion to physical
quantities is needed, like converting the internal time variable $\log
a$ to physical time $t$, or in the equation of motion in a
cosmological context. Thus, for running cosmological simulations
including dark-energy, once can rewrite the usual Hubble function
\begin{equation}
  H(a)=H_0\,\left[
    \frac{\Omega_0}{a^3}+
    \frac{1-\Omega_0-\Omega_\Lambda}{a^2}+
    \Omega_\Lambda
  \right]^{1/2}\;,
\end{equation}
for a flat cosmology with cosmological constant using
Eq.~(\ref{e:omegade}) as
\begin{eqnarray}
  H(a)&=&H_0\,\left[
    \frac{\Omega_0}{a^3}+
    \frac{1-\Omega_0-\Omega_\mathrm{Q}}{a^2}\right.\nonumber\\
  &+&\left.
    \Omega_\mathrm{Q}\,
    \exp\left(-3\int_a^1\frac{1+w(a')}{a'}\d a'\right)
  \right]\;.
\label{eq:frhs}
\end{eqnarray}
Within the new integration scheme of P-GADGET-2, this is also done
within the calculation of the drift and kick factors (Springel et al.,
in preparation).

For reasons of timing and precision, we tabulate the integral in
Eq.~(\ref{eq:frhs}) at the beginning of a run, and then interpolate
within the table during the run.

\subsection{Initial conditions}

Regarding the initial conditions, we concentrate on creating a set of
identical clusters in all cosmologies studied. Thus, we started from a
set of 17 clusters in the $\Lambda$CDM cosmology and adapted their
initial conditions to the different dark-energy cosmologies
investigated.

\begin{figure*}[ht]
  \includegraphics[width=\hsize]{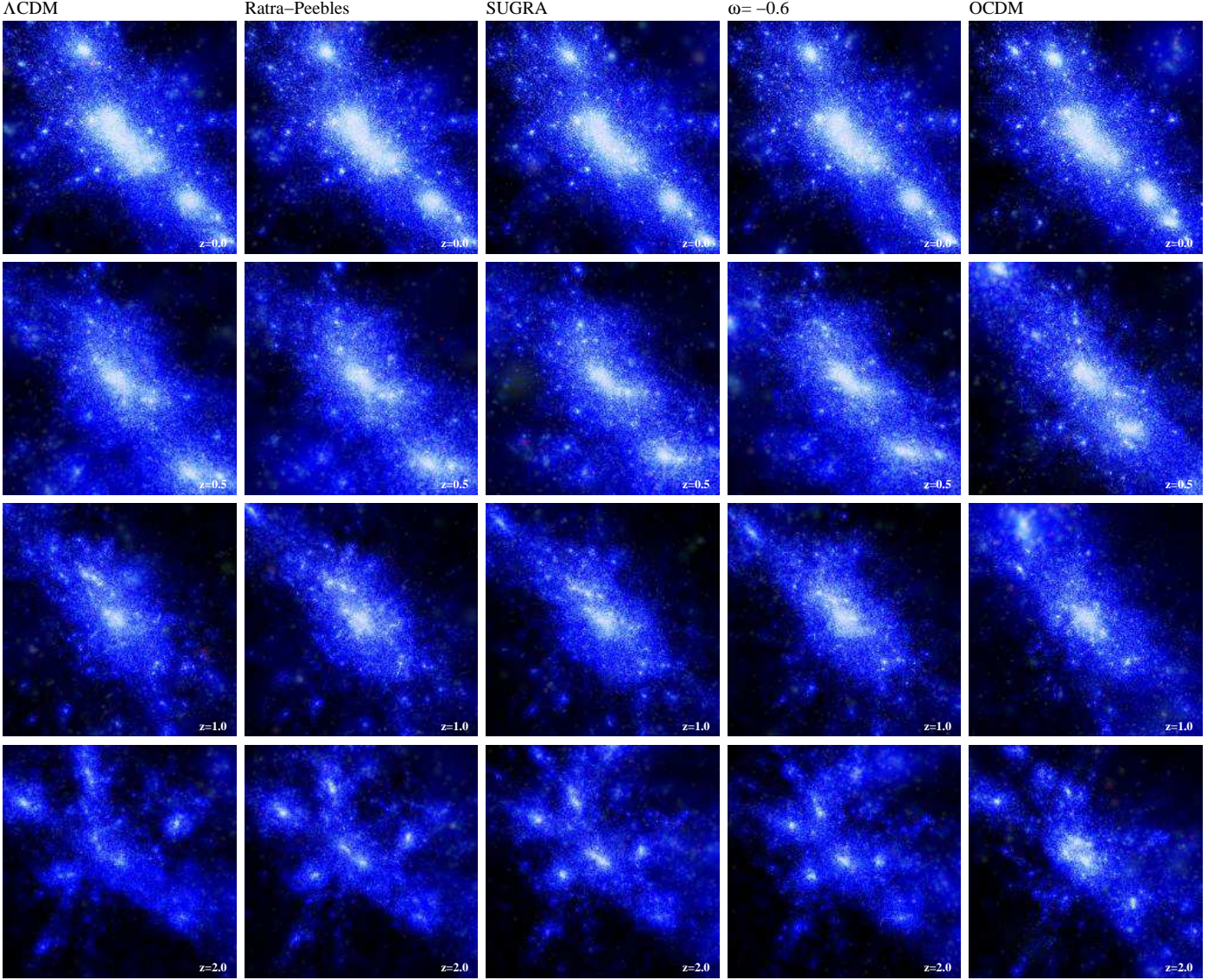}
\caption{One individual cluster is shown at different redshifts ($z=2$
  to $z=0$ in steps of $0.5$ from bottom to top) in different
  cosmologies (columns as labelled) all normalised to $\sigma_8=0.9$
  today. The panels illustrate that the clusters in different
  cosmological models arose from the same initial conditions and thus
  appear morphologically similar, but subtle differences are visible
  in detail.}
\label{fig:7}
\end{figure*}

\subsubsection{The original set of initial conditions}

For the present analysis, we use cluster models obtained using the so
called re-simulation technique. It consists of re-simulating at higher
resolution regions selected from an existing large-scale cosmological
simulation. For this work, we selected as a parent simulation an
$N$-body run with $512^3$ particles in a box of
$479\,h^{-1}\,\mathrm{kpc}$ (\citealt{YO01.1,JE01.1}). Its background
cosmological model is spatially flat with $\Omega_\mathrm{m,0}= 0.3$
and $\Omega_\Lambda=0.7$ at the final epoch, identified with redshift
zero. The Hubble constant is $h=0.7$ in units of $100\,{\rm
km\,s^{-1}\,Mpc^{-1}}$, and the power spectrum normalisation was set
to $\sigma_8=0.9$. The particle mass was
$6.8\times10^{10}\,h^{-1}\,M_\odot$, and the gravitational softening
was chosen as $30\,h^{-1}\,\mathrm{kpc}$. From the output of this
simulation at $z=0$, we randomly selected ten spherical regions of
radius between $5$ and $10\,h^{-1}\,\mathrm{Mpc}$, each containing
either one or a pair of dark matter haloes, with mass larger than
$3\times10^{14}\,M_\odot$. The total number of cluster-sized objects
turns out to be 17.

We constructed new initial conditions for each of these regions using
the ZIC software package (\emph{Zoomed Initial Conditions};
\citealt{TO97.2}). The procedure for this construction is as
follows. The initial positions of the particles in the simulation run
define a Lagrangian region. The initial density field in this
Lagrangian region is resampled by placing a larger number of particles
than were originally present; on average, the number of
high-resolution (HR), dark-matter (DM) particles is $10^6$ for our
simulations. In this way, the spatial and mass resolution can be
increased at will. The mass resolution of the re-simulation ranges
from $2\times 10^9$ to $6\times 10^9 h^{-1}\,M_\odot$ per DM particle
so as to have each cluster consists of approximately the same number
of particles. The gravitational softening is given by a
$5\,h^{-1}\,\mathrm{kpc}$ cubic spline smoothing for all HR particles.

The number of particles outside the HR Lagrangian region was reduced
by interpolating them onto a spherical grid centred on the geometrical
centre of the HR region. An angular resolution of the grid between 3
and 5 degrees in both angular directions produces on the order of
$100,000$ macro particles of varying mass and gravitational
softening. Extensive testing has shown that this number is enough to
guarantee an overall good description of the original tidal field on
large scales.

The distribution of high resolution particles in the new initial
conditions samples all the fluctuations of the matter power spectrum
realization of the original cosmological run, plus a new and
independent realization of high frequency fluctuations from the
original spectrum; in this way the power spectrum is extended up to
the Nyquist frequency of the new HR particle distribution.

The ZIC package has been widely tested and was used to produce initial
conditions for many resimulations at medium to extremely high
resolution (e.g. \citealt{TO97.2,SP01.2,YO02.1,ST02.1}).

\subsubsection{Modifying the initial conditions for dark-energy models}

Since the goal of this work is studying the evolution of the same set
of clusters in different cosmologies, we decided to adapt the initial
conditions from the $\Lambda$CDM cosmology to all the dark energy
cosmologies we wish to investigate. This is done in two steps. In the
first step, the initial redshift is adapted such that the \emph{rms}
density fluctuation amplitude today is the same in all models despite
their different dynamics. The initial redshift $z_\mathrm{ini}$ for
any model is thus implicitly determined by the ratio of linear growth
factors $D_+(z)$,
\begin{equation}
  \frac{D_+(z_\mathrm{ini})}{D_+(0)}=
  \frac{D_\mathrm{+,\Lambda CDM}(z^\mathrm{ini}_\mathrm{\Lambda CDM})}
       {D_\mathrm{+,\Lambda CDM}(0)}\;.
\end{equation}
In a second step, the peculiar velocities of the particles in their
Lagrangian initial region have to be modified such as to reflect the
change in initial redshift. At early times $\tau$, they can be
described by the \citet{ZE72.1} approximation,
\begin{equation}
   \dot{x}(t)\approx\Omega^{0.6}(t)H(t)\nabla_q\Phi(\vec q)\;,
\end{equation}
where $q$ is the Lagrangian particle position, and $\Phi(q)$ is the
velocity potential in Lagrangian space. Thus, the new velocities can
be calculated starting from the displacement field $\nabla_q\Phi(\vec
q)$ and scaling them to the new initial redshift and the new
cosmological model we wish to adapt them to. The velocities are thus
given by
\begin{equation}
  \vec v^\mathrm{ini}=\vec v_\mathrm{\Lambda CDM}^\mathrm{ini}\,
  \frac{\Omega^{0.6}(z_\mathrm{ini})\,H(z_\mathrm{ini})}
       {\Omega_\mathrm{\Lambda CDM}^{0.6}
        (z_\mathrm{\Lambda CDM}^\mathrm{ini})\,
        H_\mathrm{\Lambda CDM}^{0.6}
        (z_\mathrm{\Lambda CDM}^\mathrm{ini})}
\end{equation}
Table~\ref{tab:ini} summarises the initial redshifts and the velocity
scalings used for the different cosmologies.

\begin{table}
\caption{Parameters used for adapting the initial conditions of the
  $\Lambda$CDM model to the other cosmological models used in this
  paper. Models marked with superscript $^\dagger$ have their
  \emph{rms} fluctuation amplitude $\sigma_8$ adapted to the COBE
  normalisation, see Table~\ref{tab:1}.}
\label{tab:ini}
\begin{center}
\begin{tabular}{|l|cc|cc|cc|}
\hline
      & \multicolumn{2}{c|}{$z_\mathrm{\Lambda CDM}=35$}
      & \multicolumn{2}{c|}{$z_\mathrm{\Lambda CDM}=50$}
      & \multicolumn{2}{c|}{$z_\mathrm{\Lambda CDM}=60$} \\
model & $z_\mathrm{ini}$ & $v_\mathrm{ini}$ 
      & $z_\mathrm{ini}$ & $v_\mathrm{ini}$ 
      & $z_\mathrm{ini}$ & $v_\mathrm{ini}$ \\
\hline
$\Lambda$CDM  & 35.0 & 1.00 & 50.0 & 1.00 & 60.0 & 1.00 \\
Ratra-Peebles & 37.4 & 1.10 & 53.5 & 1.10 & 64.1 & 1.10 \\
Ratra-Peebles$^\dagger$ & 33.8 & 0.95 & 48.3 & 0.95 & 58.0 & 0.95 \\
SUGRA         & 43.0 & 1.35 & 61.7 & 1.36 & 74.1 & 1.36 \\
SUGRA$^\dagger$ & 35.9 & 1.03 & 51.5 & 1.04 & 62.0 & 1.05 \\
OCDM          & 59.4 & 2.16 & 85.1 & 2.19 & 102.2 & 2.16 \\
$w=-0.6$      & 41.8 & 1.30 & 59.7 & 1.30 & 71.6 & 1.30 \\
$w=-0.6\,^\dagger$ & 39.9 & 1.21 & 57.0 & 1.21 & 68.4 & 1.21 \\
\hline
\end{tabular}
\end{center}
\end{table}

\subsection{Post-processing}

We applied a halo finder and constructed the merger tree as described
in \citet{TO03.1}.

\subsubsection{Finding the halos}

The halo finder adopts the spherical overdensity criterion to define
collapsed structures in the simulations. This is done for each
snapshot of each resimulation, estimating the local dark matter
density at the position of each particle, $\varrho_{i;\mathrm{DM}}$,
by calculating the distance $d_{i,10}$ to the tenth closest neighbour,
and assuming $\varrho_{i;\mathrm{DM}}\propto d_{i,10}^{-3}$. We then
sort the particles by density and take as centre of the first halo the
position of the particle embedded into the highest density. Around
this centre, we grow spherical shells of matter, recording the total
mean overdensity inside the sphere as it decreases with increasing
radius. We stop the growth and cut the halo when the overdensity first
drops below 200 times the \emph{mean} (as opposed to \emph{critical})
background density, and denote the radius so defined as $r_{200}$. The
particles selected in this way belong to the same halo and are used to
compute its virial properties (mass, radius, etc.). We tag all halo
particles as \emph{engaged} in the list of sorted densities, and
selected the centre of the next halo at the position of the densest
available (unengaged) particle. We continue in this manner until all
particles are screened. We include in our catalogue only such halos
which have at least $n=10$ dark-matter particles within their virial
radius. All other particles are considered field particles.

We found that our 17 clusters contain on average $N_V\approx200,000$
dark matter particles within their virial radii. The corresponding
virial masses range between $M_V=3.1\times10^{14}$ to
$1.7\times10^{15}\,h^{-1}\,M_\odot$.

\subsubsection{Constructing merger trees}

For each of these cluster-sized halos we build a merging history tree
using the halo catalogs at all time outputs. Starting with a halo at
any given $z$, we define its progenitors at the previous output
$z+\Delta z$ to be all haloes containing at least one particle that by
$z$ will belong to the first halo. We call the main progenitor at
$z+\Delta z$ the one giving the largest mass contribution to the halo
at $z$.

\subsubsection{Fitting halo concentrations}

For determining halo concentrations and their change with time, we
only took the main progenitor of each of the 17 massive halos into
account. Using the centres from the halo finder, we construct radial
profiles by binning the particles within $r_{200}$ in logarithmic
radial bins with at least 132 particles each. For low-mass halos,
this number is reduced to be $1/20$ of the total number of particles
within $r_{200}$ to enshure enough points for the fitting procedure.
Our choice in number of particles is smaller than suggested by 
\citet{PO03.1} but still reasonable, as we are interested in the 
local density at the innermost point, and not in the enclosed
over-density. Also the global fit does not strongly depend on the
innermost data-points. We checked, that the largest contribution to the
scatter of the concentration parameter within a redshift bin comes
from the dynamical state of the halo. The changes in the concentration 
parameters inferred from the fits when changing then number of
particles within the innermost bin by a factor of five is still one
order of magnitude smaller than the intrinsic scatter due to the
dynamical state of the halo.

We then fit the NFW profile
\begin{equation}
  \rho(r) = \frac{\rho_0}{(r/r_\mathrm{s})(1+r/r_\mathrm{s})^2}
\end{equation}
\citep{NA97.1} to the binned profiles obtained. The concentration is
then defined as usual, $c=r_{200}/r_\mathrm{s}$. Note, however, that
\cite{NA97.1} defined $r_{200}$ as enclosing a mean density of 200
times the \emph{critical} rather than the \emph{mean} background
density as we do here.

\begin{figure}[ht]
  \includegraphics[width=\hsize]{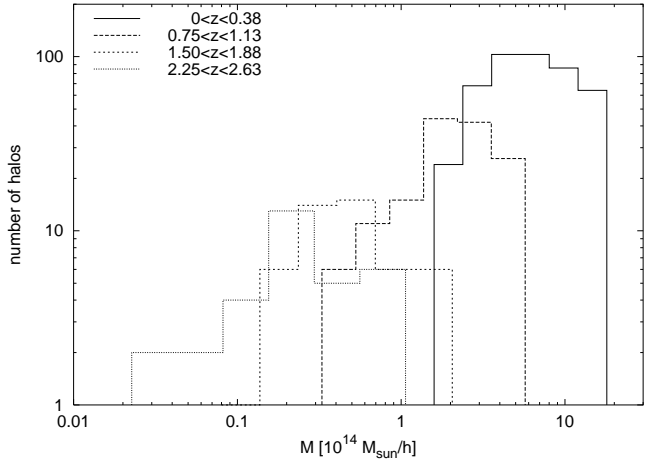}
\caption{Halo-mass histograms in the $\Lambda$CDM model for four of
  the eight equidistant redshift bins between redshifts zero and three
  introduced in Sect.~\ref{sec:4} below. The distributions in the
  missing bins fall in between those in the adjacent bins and are left
  out for clarity. As halos merge and new halos form, the histograms
  shift towards higher masses.}
\label{fig:6}
\end{figure}

\section{Halo concentrations\label{sec:4}}

Given the numerically determined halo concentrations, we can now
proceed to compare them with expectations from analytic
algorithms. Three such algorithms have been proposed:

\begin{itemize}

\item \cite{NA97.1} assign a halo formation redshift $z_\mathrm{coll}$
  requiring that half of the final halo mass $M$ be contained in
  progenitors with masses of at least $f_\mathrm{NFW}\,M$. The halo
  density scale $\rho_0$ is then assumed to be $C_\mathrm{NFW}$ times
  the mean cosmic density at $z_\mathrm{coll}$. Fitting to their
  numerically determined halo concentrations, they recommend setting
  $f_\mathrm{NFW}=0.01$ and $C_\mathrm{NFW}=3\times10^3$.

\item \cite{BU01.1} noticed that the algorithm proposed by
  \cite{NA97.1}, although well describing halo concentrations at
  redshift zero, under-predicts their decrease with increasing
  redshift. They define the collapse redshift $z_\mathrm{coll}$ by
  requiring that the non-linear mass scale at $z_\mathrm{coll}$ be a
  fraction $f_\mathrm{B}$ of the final halo mass, and assume that the
  halo concentration is a factor $C_\mathrm{B}$ times the ratio of
  scale factors at the redshift when the halo is identified and the
  collapse redshift. They recommend setting $f_\mathrm{B}=0.01$ and
  $C_\mathrm{B}=4$.

\item \cite{EK01.1} define the collapse redshift using the power
  spectrum. They require that a suitably defined amplitude of the
  dark-matter power spectrum at the mass scale $M$, linearly evolved
  to $z_\mathrm{coll}$, be equal to a constant $C_\mathrm{ENS}$, and
  find $C_\mathrm{ENS}=1/28$.

\end{itemize}

The algorithms by \cite{NA97.1} and \cite{BU01.1} involve two
parameters, a mass fraction $f_\mathrm{NFW,B}$ and a constant factor
$C_\mathrm{NFW,B}$, while \cite{EK01.1} introduce a single parameter
only.

Before we can apply these algorithms to our simulated halos, we have
to introduce a common definition of virial radii and masses. While
\cite{NA97.1} define the virial radius $r_{200}$ as enclosing an
overdensity of 200 times the \emph{critical} density of the Universe,
\cite{BU01.1} and \cite{EK01.1} refer to the \emph{mean} density
instead. Since the mean is lower than the critical density, $r_{200}$
is larger using the latter definition. Moreover, \cite{BU01.1} use the
mean overdensity within virialised spherical halos instead of the
factor 200.

In our numerical simulations, we use the radius enclosing 200 times
the \emph{mean} background density as virial radius, which we call
$r_{200}$ throughout. We adapt the numerical algorithms by
\cite{NA97.1}, \cite{BU01.1} and \cite{EK01.1} such that they use the
same definition of $r_{200}$. This requires iterative solutions
because converting masses from one definition of the virial radius to
another requires the density profile, and thus concentrations, to be
known.

\begin{figure*}[ht]
  \includegraphics[width=0.49\hsize]{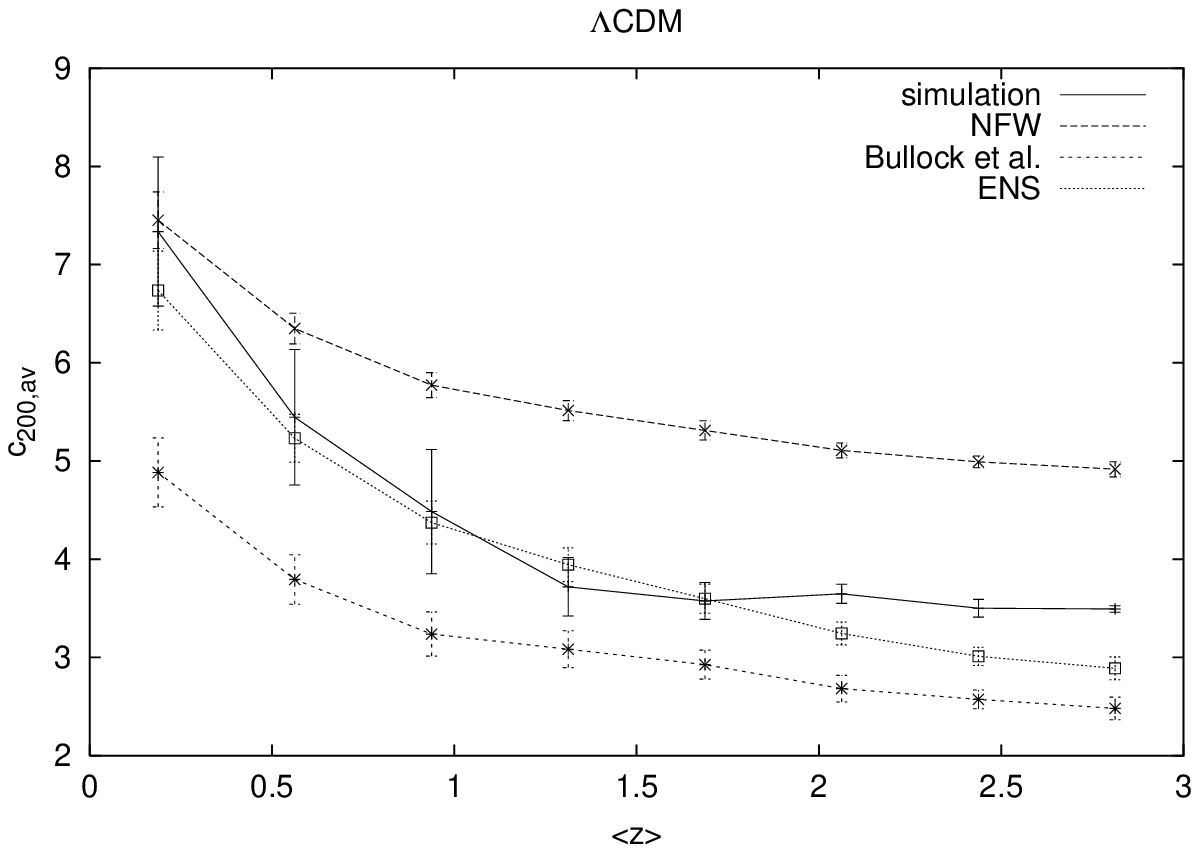}
  \includegraphics[width=0.49\hsize]{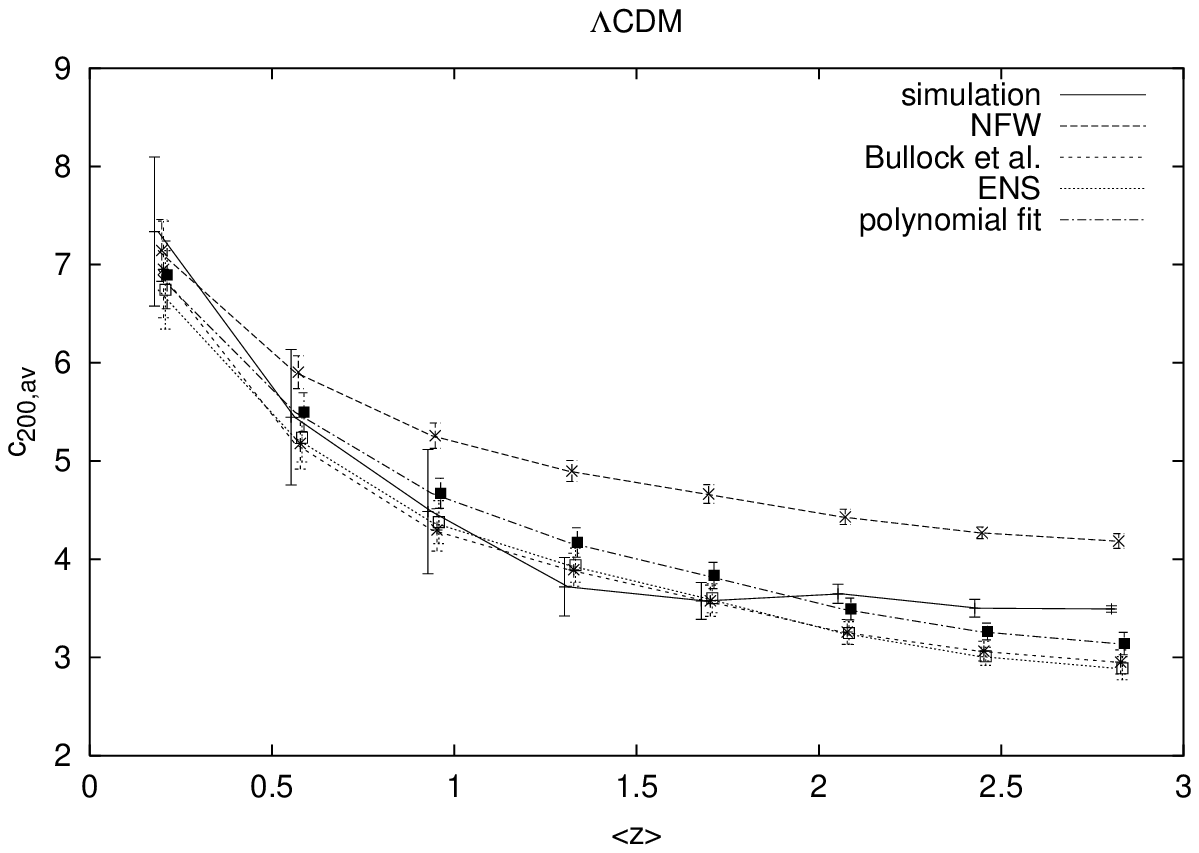}
\caption{Concentrations of halos formed in the $\Lambda$CDM model
  binned by redshift before (left panel) and after (right panel)
  determining halo-prescription parameters. While the ENS algorithm
  fits the measured halo concentrations very well, concentrations
  according to the Navarro et al. and Bullock et al. algorithms
  deviate significantly. Changing the algorithm parameters yields very
  good agreement also for the Bullock et al. algorithm, while the
  Navarro et al. algorithm has too shallow redshift evolution. The
  right panel also shows the good quality of the simple polynomial
  fit. The error bars indicate the approximate 1-$\sigma$ range of
  halo concentrations within the redshift bins. Concentrations are
  defined relative to the radius $r_{200}$ of spheres around halo
  centres containing 200 times the \emph{mean} background density.}
\label{fig:3}
\end{figure*}

Given the masses and redshifts of all individual halos, we compute the
analytically expected concentrations for each halo according to the
three algorithms listed above. We thus have for each halo four
concentration values, viz.~the three analytic expectations and the
value obtained by fitting the NFW profile to the numerically
determined density profile. We then bin halos by redshift into eight
bins between redshifts three and zero. Each redshift bin thus contains
halos with a range of masses. In each bin, we determine the median
and the 33- and 68-percentiles of the distribution of the four types
of concentration. The left panel of Fig.~\ref{fig:3} shows results for
the $\Lambda$CDM model as an example.

The solid curve shows the numerically determined halo concentrations,
with the error bars giving the 33- and 68-percentiles centred on the
median within each bin. The other curves show results obtained from
the three algorithms as indicated in the figure. Error bars attached
to those reflect the mass ranges of halos within the respective
redshift bins. While the algorithm by \cite{EK01.1} describes the
numerically fitted halo concentrations very well within the error
bars, concentrations obtained according to \cite{NA97.1} and
\cite{BU01.1} fall consistently above and below the numerical
concentrations, respectively. The curve for the NFW algorithm reflects
the earlier finding that it reproduces halo concentrations well at
redshift zero, but under-predicts their decrease with increasing
redshift. Concentrations according to \cite{BU01.1} are somewhat too
small in all redshift bins.

The error bars on the simulated curve are larger than those on the
analytically determined curves because of halo mergers, which change
halo masses much more abruptly than halo concentrations.

\begin{figure*}[ht]
  \includegraphics[width=0.49\hsize]{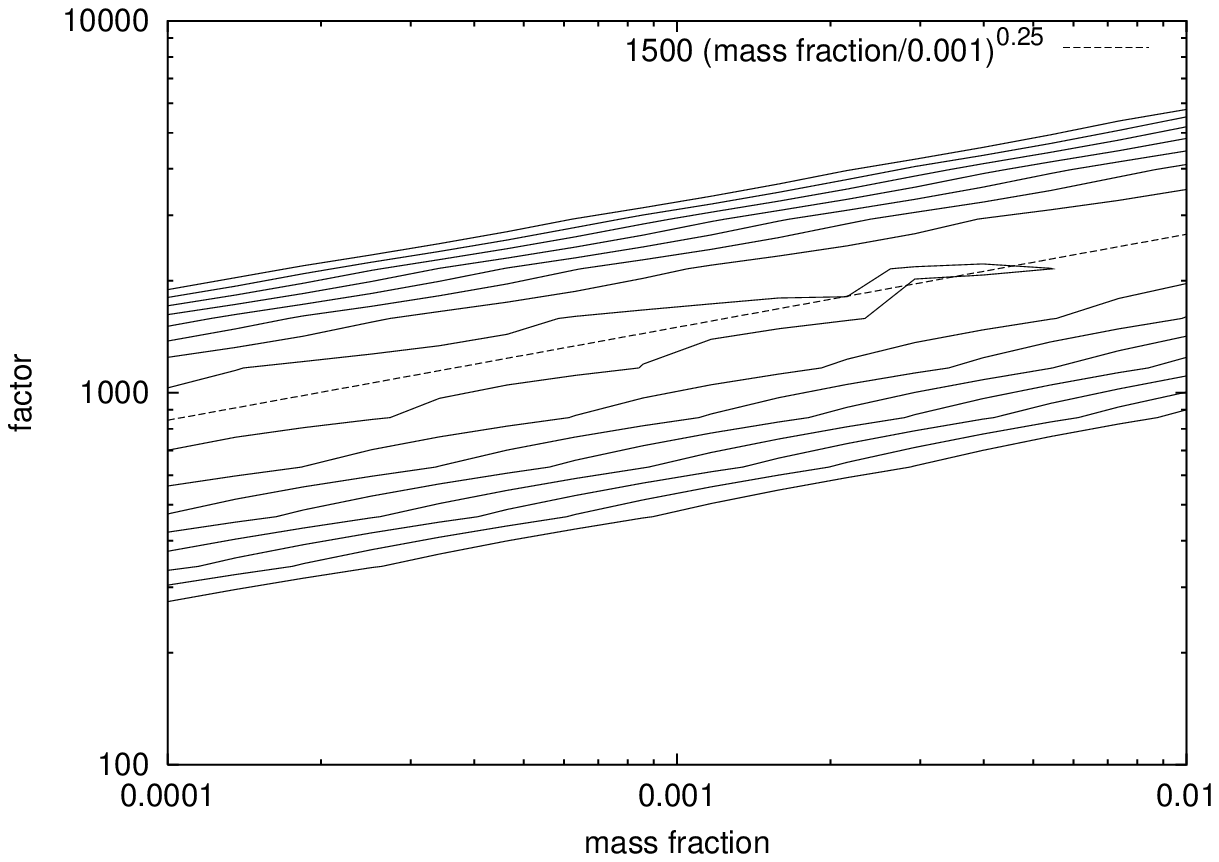}
  \includegraphics[width=0.49\hsize]{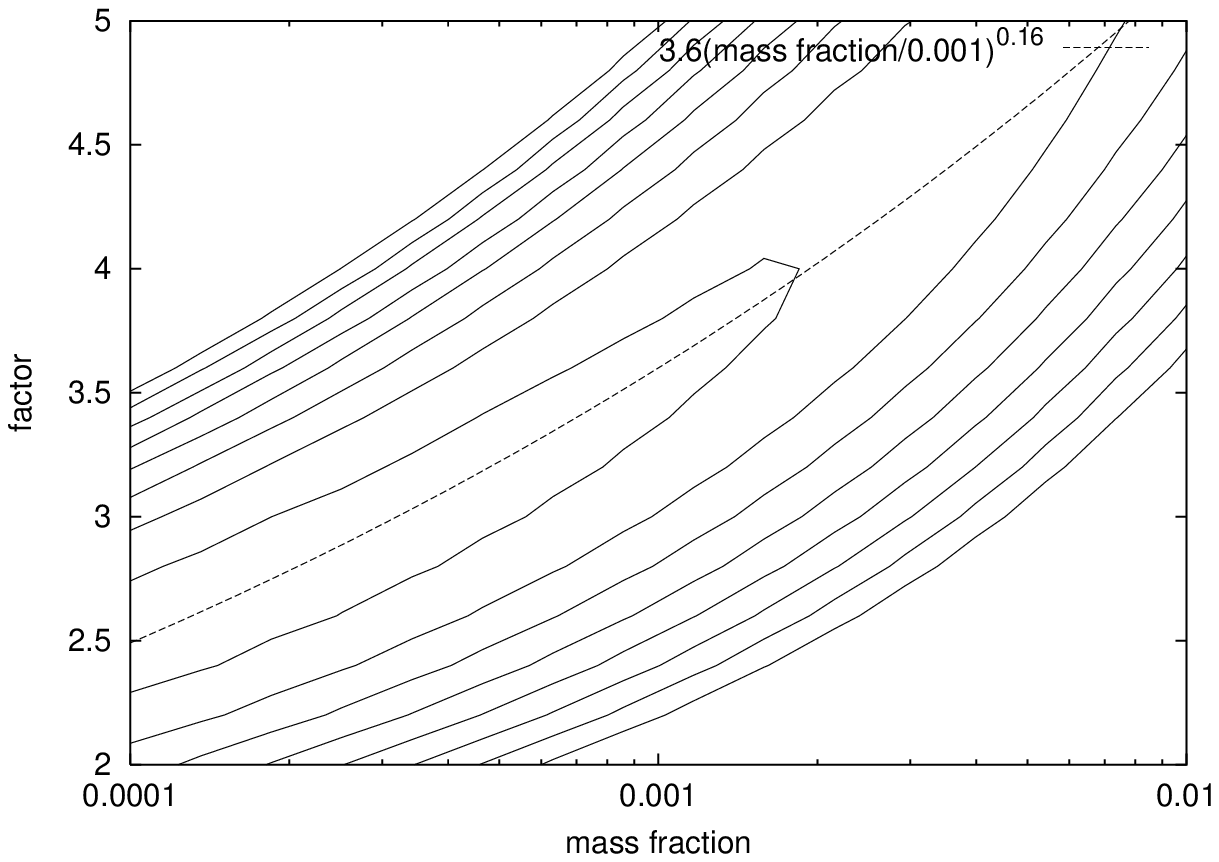}
\caption{$\chi^2$ contours in the space of parameters contained in the
  halo-concentration prescriptions by Navarro et al. (left panel) and
  Bullock et al. (right panel). Both contour plots show extended
  valleys in the $\chi^2$ surface with very flat bottoms. The dotted
  curves indicate approximate relations for the directions of these
  valleys.}
\label{fig:1}
\end{figure*}

Since all three algorithms have one or two free parameters, we now
investigate whether their agreement with the numerical results can be
improved modifying the parameters. We thus define a measure for the
quadratic deviation of analytical from numerical halo concentrations,
\begin{equation}
  \chi^2=\sum_\mathrm{halos}\left[\left(
    c_\mathrm{analytic}-c_\mathrm{numerical}
  \right)^2\right]\;,
\label{eq:1}
\end{equation}
and minimise $\chi^2$ varying the parameter(s). Figure~\ref{fig:1}
shows $\chi^2$ contours obtained for the $\Lambda$CDM model from the
algorithms by \cite{NA97.1} (left panel) and \cite{BU01.1} (right
panel). Both algorithms have two parameters, a mass fraction and a
factor, which are shown along the horizontal and vertical axes,
respectively.

Both contour plots share the feature of having a shallow, degenerate
valley along which the mass fraction changes by orders of magnitude,
while the factor changes very little. Approximately, these valleys
follow the relations indicated by dashed lines in the two panels of
Fig.~\ref{fig:1}. This near-insensitivity to the exact mass fraction
allows fixing it at $f_\mathrm{NFW,B}=0.001$, say, and then
determining the $\chi^2$ minimum along the perpendicular axis only.

\begin{figure}[ht]
  \includegraphics[width=\hsize]{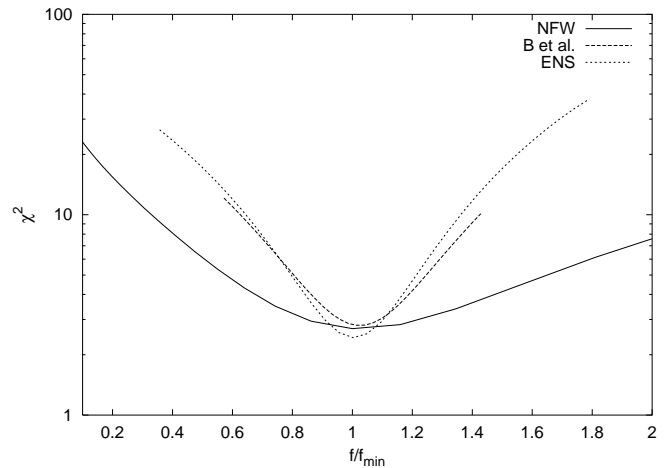}
\caption{Cuts through the $\chi^2$ contours at constant mass fraction
  for three different halo-concentration prescriptions.}
\label{fig:2}
\end{figure}

Figure~\ref{fig:2} shows cuts through the contour plots in
Fig.~\ref{fig:1}, and also $\chi^2$ according to the one-parameter
algorithm proposed by \cite{EK01.1}. The abscissa shows the factors
$C_\mathrm{NFW,B}$ divided by their values at the $\chi^2$ minima for
fixed $f_\mathrm{NFW,B}=0.001$, i.e.~$C_\mathrm{NFW}=1500$ and
$C_\mathrm{B}=3.5$, while $C_\mathrm{ENS}$ is divided by the
originally proposed value, i.e.~$C_\mathrm{ENS}=1/28$. The curves show
pronounced minima for the algorithms proposed by \cite{BU01.1} and
\cite{EK01.1}, and a very shallow minimum for the \cite{NA97.1}
algorithm.

Using these parameters, we now return to the comparing the numerically
determined and analytically expected halo concentrations and find the
results shown in the right panel of Fig.~\ref{fig:3}. The agreement
between the numerical results and the \cite{BU01.1} concentrations has
improved substantially, while the \cite{NA97.1} algorithm still
predicts too high halo concentrations at moderate and high redshifts.

\begin{figure*}[ht]
  \includegraphics[width=0.49\hsize]{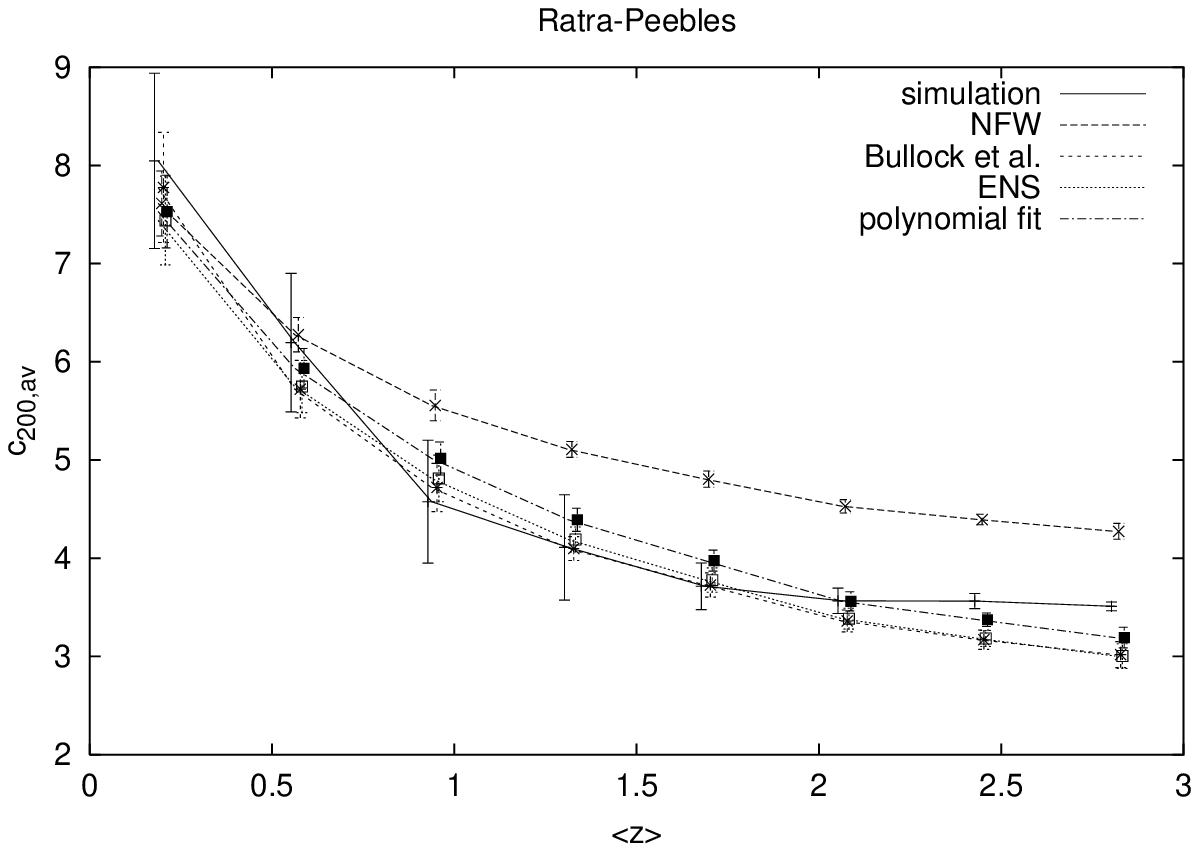}
  \includegraphics[width=0.49\hsize]{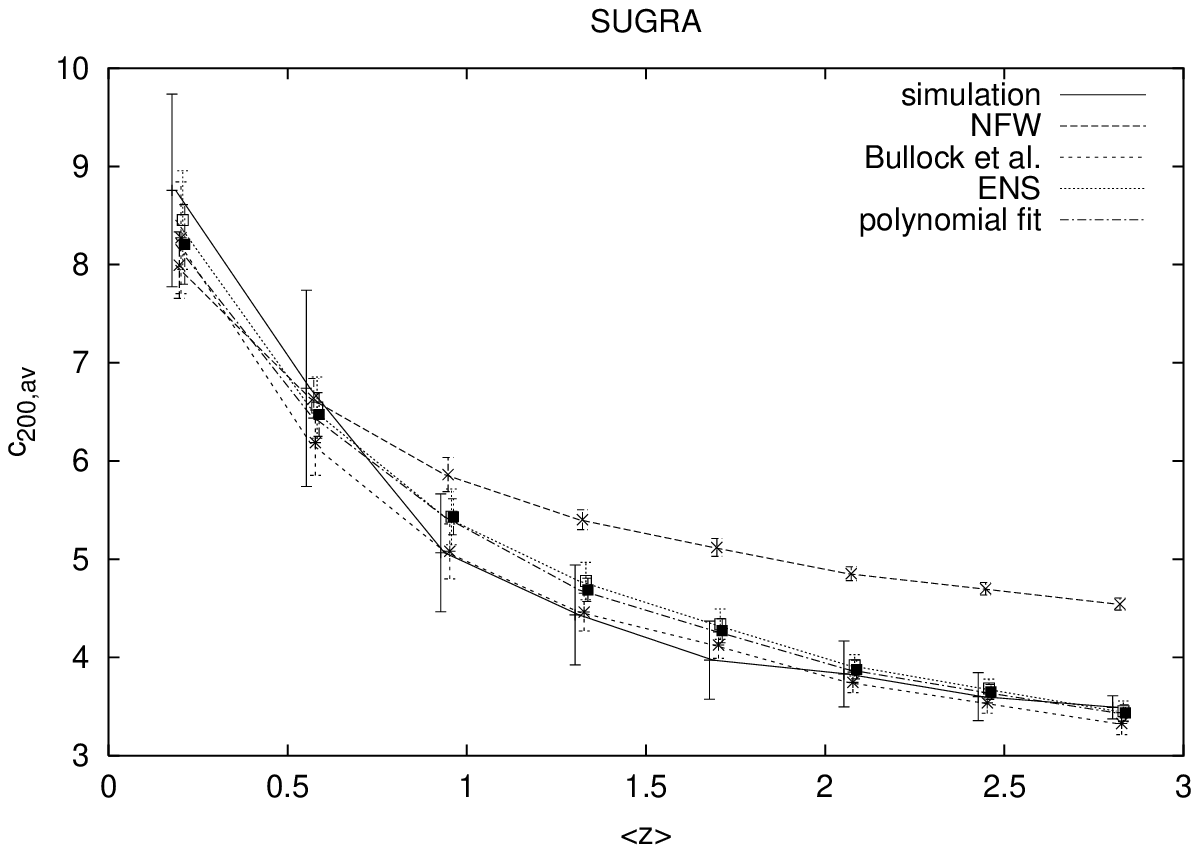}
\caption{Examples for the agreement between numerically simulated and
  analytically expected halo concentrations for two dark-energy
  models, Ratra-Peebles (left panel) and SUGRA (right panel). The
  agreement is very good, except for the Navarro et al. prescription
  whose redshift evolution is too shallow.}
\label{fig:4}
\end{figure*}

We repeated this procedure for all cosmological models used and found
that the algorithm by \cite{BU01.1} reproduces the numerical
concentrations well in all of them if its parameters are modified to
$f_\mathrm{B}=0.001$ and $C_\mathrm{B}=3.5$, while the algorithm by
\cite{EK01.1} performs very well throughout with the factor
$C_\mathrm{ENS}=1/28$ which was originally proposed. The agreement
achieved for dark-energy models is illustrated for the Ratra-Peebles
and SUGRA models in the left and right panels of Fig.~\ref{fig:4},
respectively.

Our first conclusion is thus that the halo-concentration algorithms
proposed by \cite{BU01.1} performs very well also in dark-energy
cosmologies, provided its two parameters are modified, while the
algorithm by \cite{EK01.1} does not require any adaptation. The
algorithm originally proposed by \cite{NA97.1} has the same weakness
in dark-energy as in $\Lambda$CDM models of under-predicting the
redshift evolution. This can be remedied to some degree, but not
removed, by modifying its two parameters.

These algorithms have the advantage of relating halo properties to the
physical mechanism of halo formation. The gentle changes of halo
concentrations with mass and redshift suggest, however, that they can
be described in a much more simplified fashion by a fitting
formula. Since concentrations of halos with fixed mass increase with
decreasing redshift as the background density decreases, we attempt
fitting the two-parameter functional relationship
\begin{equation}
  c(M,z)=\frac{c_0}{1+z}\,
  \left(\frac{M}{10^{14}\,h^{-1}\,M_\odot}\right)^\alpha
\label{eq:2}
\end{equation}
to the numerically determined halo concentrations. Best-fitting values
for $c_0$ and $\alpha$ are listed for all eight cosmological models in
Tab.~\ref{tab:1}. The uncertainties of $c_0$ and $\alpha$ given there
are 1-$\sigma$ bootstrap errors obtained from 1024 bootstrap halo
samples. Although Eq.~(\ref{eq:2}) does not reflect the physics of
halo formation, it is useful for conveniently summarising the
numerical results.

\begin{table}
\caption{Parameters $c_0$ and $\alpha$ for the different cosmological
  models}
\label{tab:1}
\begin{center}
\begin{tabular}{|l|rrr|}
\hline
model & $\sigma_8$ & $c_0$ & $\alpha$ \\
\hline
$\Lambda$CDM  & 0.90 & $ 9.59\pm0.07$ & $-0.102\pm0.004$ \\
Ratra-Peebles & 0.90 & $10.20\pm0.07$ & $-0.094\pm0.005$ \\
Ratra-Peebles & 0.82 & $ 9.30\pm0.06$ & $-0.108\pm0.005$ \\
SUGRA         & 0.90 & $11.15\pm0.09$ & $-0.094\pm0.006$ \\
SUGRA         & 0.76 & $ 9.46\pm0.07$ & $-0.099\pm0.005$ \\
OCDM          & 0.90 & $14.29\pm0.13$ & $-0.089\pm0.006$ \\
$w=-0.6$      & 0.90 & $11.32\pm0.09$ & $-0.092\pm0.005$ \\
$w=-0.6$      & 0.86 & $10.44\pm0.08$ & $-0.066\pm0.005$ \\
\hline
\end{tabular}
\end{center}
\end{table}

The agreement between the numerically determined concentrations and
those obtained from the fitting formula (\ref{eq:2}) are illustrated
for $\Lambda$CDM in the right panel of Fig.~\ref{fig:3} and for two
dark-energy models in both panels of Fig.~\ref{fig:4}. While the
constant factor $c_0$ changes quite appreciably across cosmological
models, the exponent $\alpha\approx-0.1$ is approximately constant for
all models, except perhaps for the dark-energy model with constant
$w=-0.6$ and reduced $\sigma_8$. Recently, \cite{ZH03.1} found that
the dependence of halo concentrations on halo mass becomes shallower
with increasing redshift. Since we are focussing on how halo
properties vary across cosmologies, our halo sample is currently too
small for confirming this result.

\begin{figure}[ht]
  \includegraphics[width=\hsize]{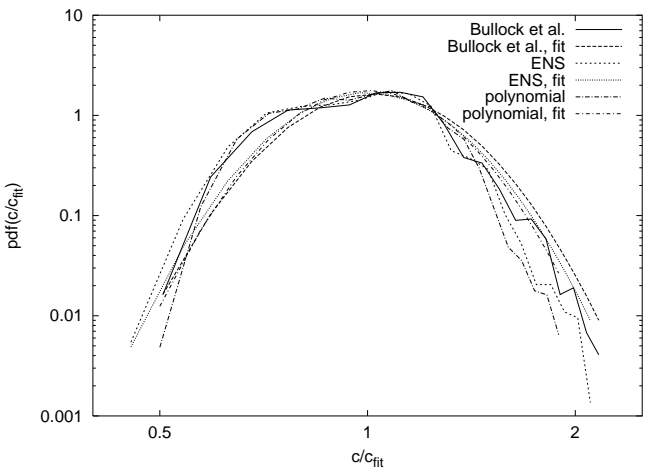}
\caption{Normalised halo-concentration distributions across all
  cosmological models for three halo-concentration algorithms. The
  curves are very close to log-normal distributions whose standard
  deviations are almost independent of the cosmological model.}
\label{fig:5}
\end{figure}

Remarkably, halo concentrations are distributed about their mean
values $\bar{c}$ in a way which is virtually independent of the
cosmological model. Figure~\ref{fig:5} shows the distribution of
$c/\bar{c}$ for all eight cosmologies, where separate curves are shown
for $\bar{c}$ determined according to the algorithms by \cite{BU01.1},
\cite{EK01.1} and the fitting formula (\ref{eq:2}).

These curves closely resemble log-normal distributions, as found
earlier in $\Lambda$CDM and OCDM cosmologies
(e.g.~\citealt{JI00.1,BU01.1}) . We thus attempt fitting them with
distributions of the form
\begin{equation}
  p(c)=A\,\exp\left[
    -\frac{(\ln c-\ln\bar{c}-\ln\mu)^2}{2\sigma^2}
  \right]\;.
\label{eq:3}
\end{equation}
The amplitude $A$ is fixed by normalisation, and the parameters $\mu$
and $\sigma$ are determined from fitting to the measured
distributions, using $\bar{c}$ as obtained from the different
algorithms and the polynomial of (\ref{eq:2}).

Any deviation of $\mu$ from unity indicates that the respective
distribution is not centred on $\bar{c}$. Since the algorithms and
Eq.~(\ref{eq:2}) were adapted such as to optimise the agreement
between $\bar{c}$ and the measured mean values, $\mu$ is determined
for cross-checking only. Table~\ref{tab:2} shows the results for the
three algorithms and the polynomial (\ref{eq:2}).

\begin{table*}
\caption{Means and standard deviations of log-normal distributions
  fitted to normalised halo concentration distributions for four
  different halo-concentration prescriptions in the eight different
  cosmological models}
\label{tab:2}
\begin{center}
\begin{tabular}{|l|rrrr|rrrr|}
\hline
model & \multicolumn{4}{c|}{mean} & \multicolumn{4}{c|}{$\sigma$} \\
      & NFW & B & ENS & poly & NFW & B & ENS & poly \\
\hline
$\Lambda$CDM  & 0.91 & 1.04 & 1.05 & 1.00 & 0.23 & 0.22 & 0.22 & 0.21 \\
Ratra-Peebles & 0.93 & 1.01 & 1.03 & 1.00 & 0.23 & 0.22 & 0.22 & 0.21 \\
Ratra-Peebles & 0.90 & 1.00 & 1.02 & 1.00 & 0.23 & 0.22 & 0.22 & 0.22 \\
SUGRA         & 0.95 & 1.02 & 0.98 & 1.00 & 0.24 & 0.23 & 0.23 & 0.22 \\
SUGRA         & 0.92 & 1.02 & 0.98 & 1.00 & 0.25 & 0.24 & 0.24 & 0.23 \\
OCDM          & 0.99 & 1.10 & 0.91 & 1.00 & 0.25 & 0.24 & 0.23 & 0.23 \\
$w=-0.6$      & 0.95 & 1.00 & 1.01 & 1.00 & 0.23 & 0.23 & 0.22 & 0.22 \\
$w=-0.6$      & 0.92 & 0.99 & 1.00 & 1.00 & 0.24 & 0.23 & 0.23 & 0.22 \\
combined      & 0.93 & 1.02 & 1.00 & 1.00 & 0.24 & 0.23 & 0.23 & 0.22 \\
\hline
\end{tabular}
\end{center}
\end{table*}

The table shows that the mean $\mu$ scatters slightly about unity for
all algorithms except for \cite{NA97.1}, for which it is consistently
below unity. This reflects the well-known inability of that algorithm
to model the redshift evolution correctly: It over-predicts
concentrations $\bar{c}$ at redshifts above unity, and thus shifts the
distribution of $c/\bar{c}$ systematically to values below
unity. Perhaps more surprising is that the standard deviation of the
log-normal distribution is quite independent of the cosmological model
and of the algorithm used for describing the halo concentrations,
$\sigma\approx0.22$. This result agrees well with the scatter found in
a variety of CDM models with and without cosmological constant
\citep{JI00.1,BU01.1}.

\section{Dependence on $w(z)$\label{dow}}

\begin{figure}
\includegraphics[width=\hsize]{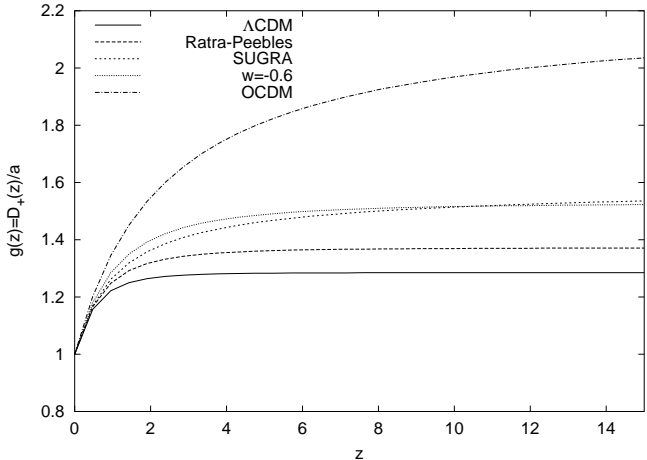}
\caption{Linear density perturbation growth dynamics in the different
  cosmologies considered.}
\label{D_over_a_DE_LAMBDA_OPEN_RP_SUGRA}
\end{figure}

In this section, we focus on the models with fixed $\sigma_8=0.9$, in
order to study the dependence of the halo concentration on the
behaviour of $w(z)$. As we saw in the previous section, $c_0$ for fixed
normalisation $\sigma_8$ increases for the following order of
cosmologies: $\Lambda$CDM, RP, SUGRA, DECDM, and OCDM. This is not
accidental, but expresses one of our major results. As we stressed in
the introduction, the different candidates for explaining the dark
energy and cosmological acceleration are different in the redshift
dependence of their equation of state. The models we study here, in
particular the RP and SUGRA quintessence scenarios, well represent
this issue, having the same $w_0$ but a markedly different $w(z)$, as
is obvious from Fig.~\ref{w_z_RP_SUGRA}. Thus, it is worthwhile
studying the difference between these models in a phenomenological
context. In Fig.~\ref{D_over_a_DE_LAMBDA_OPEN_RP_SUGRA}, we plot the
linear perturbation growth factor divided by the scale factor,
$g(z)\equiv D_+(z)/a(z)$, for the cosmologies studied here. $D_+$ is
the growing solution of the general linear perturbation equation
\begin{equation}
  \ddot\delta+2\frac{\dot a}{a}\dot\delta-4\pi G\rho\delta=0
\label{D+}
\end{equation}
which can, for $w=-1$ and $w=-1/3$, be integrated once to yield
$D_+\propto\dot{a} a^{-1}\int_0^a \dot{a}^{-3} \d a \;$.
We normalise $D_+$ to unity at present.

The asymptotic behaviour $D_+\propto a$ for $a\to0$ means that the
growth factor at early times in all cosmologies converges to the
behaviour in a flat CDM Einstein-de Sitter case, with $\Omega_0=1$. On
the other hand, the growth factor normalised by the scale factor,
$g(z)$, behaves differently for the different cosmologies. Most
importantly, it turns out to reflect the behaviour of $c_0$ in the
different cosmologies studied here above.

The curves in Fig.~\ref{D_over_a_DE_LAMBDA_OPEN_RP_SUGRA} display
$g(z)$ for the different cosmological models, normalised to unity at
present: the higher $g(z)$ is at a given redshift $z$, the higher was
the perturbation amplitude at that epoch, and the earlier was the
epoch when structures formed \citep{BA02.1}. Since at high redshifts
all cosmologies approach the flat CDM Einstein-de Sitter limit in
which $D_+\propto a$, the structures acquire their fluctuation
amplitude at redshifts determined by the ratio of the asymptotic
values of $g(z)$ as shown in
Fig.~\ref{D_over_a_DE_LAMBDA_OPEN_RP_SUGRA}.

It is thus natural to expect that the concentration parameters $c_0$
of our halos are affected similarly in the different cosmologies,
since they parametrise the central density contrast of the
halos. This is indeed what we find, as Tab.~\ref{cd} shows.

\begin{table}
\label{cd}
\caption{Concentration and perturbation asymptotic growth factor in
  the different cosmologies.}
\begin{center}
\begin{tabular}{|l|l|l|l|}
\hline 
cosmology & $\displaystyle\frac{c_0}
             {c_0^\mathrm{\Lambda CDM}}$ &
            $\displaystyle\frac{D_+(\infty)}
             {D_+^\mathrm{\Lambda CDM}(\infty)}$ &
            $\displaystyle\frac{D_+(z_\mathrm{coll})}
             {D_+^\mathrm{\Lambda CDM}(z_\mathrm{coll})}$ \\
\hline
$\mathrm{\Lambda CDM}$ & 1 & 1 & 1 \\
RP    & $1.06\pm0.011$ & 1.07 & $1.07\pm0.0003$ \\
SUGRA & $1.16\pm0.013$ & 1.22 & $1.19\pm0.004$  \\
DECDM & $1.18\pm0.013$ & 1.19 & $1.18\pm0.001$  \\
OCDM  & $1.49\pm0.018$ & 1.64 & $1.61\pm0.01$   \\
\hline
\end{tabular}
\end{center}
\end{table}

The table has four columns. The first abbreviates the cosmological
model. The second gives the $c_0$ parameters relative to
$c_0^\mathrm{\Lambda CDM}$ for the $\Lambda$CDM model, with error bars
obtained from the bootstrap errors of $c_0$ as given in
Tab.~\ref{tab:1}. The third column shows the linear growth factors at
infinity relative to the $\Lambda$CDM model, and the fourth column
gives the same ratio taken at the average collapse redshifts of our
numerically simulated halos. Error bars on the values in the fourth
column are due to the mass range of halos which implies a range of
collapse redshifts.

The collapse redshifts were obtained using the prescription by
\cite{EK01.1} because their algorithm for computing halo
concentrations turned out to reflect our numerical results best. We
draw three main conclusions from the results in Tab.~\ref{cd}. First,
the concentration parameters $c_0$ relative to $\Lambda$CDM are quite
close to the linear growth factors at high redshift relative to
$\Lambda$CDM, although the match is not perfect. We may be affected by
cosmic variance because the number of halos per cosmology is
relatively small. Second, using different definitions of the collapse
redshift yields different ratios of the linear growth factor. In
particular, such definitions according to which the collapse redshifts
are small fail in reproducing the trend in $c_0$ with different
cosmological models. Third, since the asymptotic ratio of growth
factors towards infinite redshift is rather close to the ratios of the
$c_0$ parameters, the final halo concentrations are apparently
determined at very high redshift already even though the halos are
quite massive and thus completely assembled late in cosmic history.

We thus propose to interpret the $c_0$ parameter as composed of a
factor valid for to the $\Lambda$CDM cosmology, multiplied by a
correction which takes into account the asymptotic behaviour of the
linear growth factors at high redshift for the given dark-energy
cosmology,
\begin{equation}
  c_0\rightarrow c_0^\mathrm{\Lambda CDM}\cdot
  \frac{D_+(z_\mathrm{coll})}
       {D_+^\mathrm{\Lambda CDM}(z_\mathrm{coll})}\;.
\label{c_0_corrected}
\end{equation}
This result is particularly important since it demonstrates that
quintessence cosmologies having the same equation of state at present
but a different redshift behaviour can produce relevant and
predictable differences in the central regions of clusters.

\section{Summary}

We used cosmological numerical simulations to study the concentration
parameters of massive dark-matter halos in cosmological models with
dark energy, and compared them to results obtained in
cosmological-constant ($\Lambda$CDM) and low-density, open (OCDM)
models.

We investigate three different dark-energy cosmologies. One has a
constant ratio $w=-0.6$ between pressure and density of the dark
energy, the other two have equations of state which change over
time. The Ratra-Peebles \citep{PE03.1} model describes the dark-energy
scalar field with an inverse power-law potential, the SUGRA
\citep{BR00.2} model modifies the power-law potential with an
exponential factor.

Dark energy with non-constant density, and more generally with
non-constant equation of state, modifies terms in Friedmann's
equation. We modified and extended the proprietary GADGET-2 code
accordingly, which is an extension of the publically available GADGET
code \citep{SP01.1}.

Cluster-sized halos were selected from a large cosmological simulation
volume and re-simulated at substantially higher resolution. Since the
large-scale simulation was originally prepared for a
cosmological-constant ($\Lambda$CDM) cosmology, the initial conditions
used for re-simulation needed to be carefully adapted in two ways to
the other cosmological models. First, microwave-background (CMB)
observations on large angular scales constrain the amplitude of
dark-matter fluctuations, conventionally expressed by the \emph{rms}
fluctuation level $\sigma_8$ within spheres of $8\,h^{-1}\mathrm{Mpc}$
radius. Due to the enhanced dynamics of the gravitational potential in
dark-energy cosmologies, the integrated Sachs-Wolfe effect increases
and thus the amount of primordial fluctuations on large scales needs
to be reduced in order to remain compatible with observed CMB
fluctuations on large angular scales. This reduction of $\sigma_8$
requires the redshift of the initial conditions to be adapted. Second,
the particle velocities in the initial data need to be scaled
according to the modified cosmological dynamics in dark-energy
cosmologies.

It should be noted that the correction of $\sigma_8$ tries to adapt
the density-fluctuation level in the simulations to CMB observations
on large angular scales, while we are interested in the formation of
cluster-sized halos on much smaller scales. How cluster-sized
fluctuations are related to large-scale fluctuations depends sensitively
on the primordial slope of the dark-matter power spectrum, which we
assume to unity here.

We ran simulations for two sets of normalisations. As $\sigma_8=0.9$
for the original $\Lambda$CDM model, we first ran simulations with the
same $\sigma_8$ for the OCDM and the three dark-energy models in order
to arrive at cluster samples which could most directly be compared to
each other. We then ran three additional sets of simulations after
adapting the normalisation of the dark-energy models to account for
the enhanced integrated Sachs-Wolfe effect. There are thus in total
eight sets of simulations. Each of those contained 17 massive halos
re-simulated at high resolution.

Concentrations for these halos were determined by fitting the NFW
density profile at 78 redshifts between $z=5.8$ and $z=0$. Comparing
these fitted values to analytical expectations on halo concentrations,
we found the following results:

\begin{itemize}

\item Algorithms for predicting halo concentrations from their mass
  and redshift as proposed by \cite{NA97.1}, \cite{BU01.1} and
  \cite{EK01.1} perform differently in reproducing the concentrations
  measured in the simulations. The algorithms by \cite{NA97.1} and
  \cite{BU01.1} have two parameters each, a fraction of the final halo
  mass required in defining the collapse redshift, and a factor
  relating core halo densities to the (mean or critical) background
  density at the collapse time. While the algorithm by \cite{BU01.1}
  can be adapted modifying their originally proposed parameters, the
  algorithm by \cite{NA97.1} consistently predicts a too shallow
  redshift evolution of halo concentrations. The algorithm by
  \cite{EK01.1} succeeds very well in reproducing halo concentrations
  in all cosmologies studied without adaptation.

\item In all cosmologies studied, the halo concentrations permit the
  simple polynomial fit of Eq.~(\ref{eq:2}). While the mass dependence
  is shallow and quite independent of cosmology, the normalisation
  $c_0$ of the concentrations depends quite sensitively on the
  cosmological model. Halo concentrations are higher in models with
  time-varying dark energy than in cosmological-constant models. These
  findings confirm numerically what we had suspected in an earlier
  analytic study \citep{BA02.1}.

\item Once normalised to their analytically expected mean values, halo
  concentrations closely follow a log-normal distribution with a
  standard deviation of $\sigma\approx0.22$, almost independent of the
  cosmological model.

\item The change of the concentration normalisation $c_0$ with
  cosmology relative to the $\Lambda$CDM model is well represented by
  the respective ratio of linear growth factors at halo-collapse
  redshifts, e.g.~as defined in the halo-concentration algorithm by
  \cite{EK01.1}. Since this definition yields fairly high redshifts,
  this result seems to indicate that halo properties are defined very
  early in cosmic history, in any case much earlier than the
  relatively late cosmological epoch when cluster-sized halos accrete
  most of their mass.

\end{itemize}

Similar findings on the log-normal distribution of halo concentrations
and its standard deviation have be achieved before for CDM models with
and without cosmological constant (see, e.g.,
\citealt{JI00.1,BU01.1}). Our study extends these investigations to
dark-energy models. Our two most important conclusions are that (1)
mean halo concentrations and their distribution about those mean
values are as in low-density open and cosmological-constant models
except that (2) halos are systematically more concentrated in
dark-energy models in a way closely reflecting the linear
density-fluctuation amplitude at high redshifts.

\acknowledgements{We are deeply indebted to Volker Springel for
providing access to, and support in using and modifying P-GADGET-2
prior to release. Simon White's comments helped improving the paper
substantially. The simulations were carried out on the IBM-SP4 machine
at the ``Centro Interuniversitario del Nord-Est per il Calcolo
Elettronico'' (CINECA, Bologna), with CPU time assigned under an
INAF-CINECA grant. K.~Dolag acknowledges support by a Marie Curie
Fellowship of the European Community program ``Human Potential´´ under
contract number MCFI-2001-01227.}

\bibliography{../TeXMacro/master}
\bibliographystyle{../TeXMacro/aa}

\end{document}